# Hybrid TPMS-based Designs for Anisotropic Acoustic Metamaterials: Numerical Simulation and Topology Optimization

Mohamed Shendy[1*], Nima Maftoon[2], Armaghan Salehian[1]

[1]Department of Mechanical and Mechatronics Engineering, University of Waterloo, Waterloo, Ontario, Canada

[2]Department of Systems Design Engineering, University of Waterloo, Waterloo, Ontario, Canada

**Abstract**

Anisotropic acoustic metamaterials have received significant scholarly attention in recent years due to their capacity to manipulate wave propagation across various directions. This property is integral to applications involving directional wave guiding. Nonetheless, previously proposed anisotropic acoustic metamaterials exhibited commonalities in bandgaps along x and y periodicities, limiting their efficacy for such applications. Therefore, this work introduces hybrid Triply Periodic Minimal Surfaces (TPMSs)-based anisotropic acoustic metamaterials that manifest abundant bandgap and passband characteristics along the x and y periodicities. Four design families were considered: Primitive-Primitive sheet-based, Gyroid-Gyroid sheet-based, Diamond-Diamond sheet-based, and Nevious-Nevious sheet-based. A computationally efficient exhaustive search was employed to identify optimal hybrid metamaterial designs for various desired audible frequency ranges (i.e., ranging from 20 Hz to 20 kHz). In total, 392,178 designs were assessed in this search. The optimal designs demonstrated pronounced bandgap and passband characteristics across both periodicities, thereby positioning them as promising candidates for directional wave guiding applications. For instance, the optimal Nevious-Nevious sheet-based unit cell design achieved a minimum coverage of 62.90% for bandgaps and passbands within the frequency range of 20 Hz to 5 kHz. Following the identification of optimal unit cell designs, the acoustic pressure responses of hybrid anisotropic acoustic metamaterials, constructed from repeated optimal unit cells, were computed. This involved considering two actuation scenarios, which pertain to exciting the system along both x and y periodicities, revealing responses in alignment with optimized band characteristics.

[*]Corresponding author: mshendy@uwaterloo.ca

# 1. Introduction

The discipline of guiding acoustic waves has garnered significant attention in numerous engineering applications in recent years. Domains such as telecommunications [1,2], signal processing [3,4], and noise filtering [5,6] necessitate devices capable of manipulating waves across various directions. For example, in noise filtering applications, to achieve precise measurements with a microphone, a wave-guiding device is essential to permit the passage of desired sound waves from a single direction while filtering out noise emanating from alternative directions [7].

Recently, there has been a significant focus on artificially engineered structures known as Acoustic Metamaterials in the context of guiding acoustic waves. These metamaterials are characterized by their periodic structures, which enable them to demonstrate bandgaps and passbands within the acoustic spectrum. Bandgaps refer to the frequency ranges where the metamaterial impedes wave transmission, while passbands denote the frequency ranges where the metamaterial permits wave propagation [8]. By exhibiting bandgaps and passbands, these metamaterials can effectively manipulate acoustic waves, thereby making them highly suitable for applications related to wave guidance. Notably, both bandgaps and passbands exhibit sensitivity to the structural configuration of acoustic metamaterials, thereby rendering it a topology optimization problem. This necessitates that designers evaluate a variety of geometries to develop an acoustic metamaterial design capable of manifesting bandgaps and passbands within specified frequency ranges. Earlier efforts associated with acoustic metamaterials predominantly aimed at demonstrating the sensitivity of bandgaps and passbands to the metamaterials' topology rather than optimizing the designs. The designs pertaining to acoustic metamaterials can be broadly categorized into two distinct families. The first design family conceptualizes acoustic metamaterials by varying shells and slit distributions, culminating in the formation of distinct bandgaps and passbands [9–15]. The aforementioned literature found that acoustic metamaterials characterized by narrow slits tend to develop wider bandgaps and narrower passbands. The second design family approaches the design of acoustic metamaterials through constructing structures predicated on level curve surfaces, particularly, Triply Periodic Minimal Surfaces (TPMSs), and altering their porosity rates, commonly referred to as relative density [16–20]. Modifying the relative density generates varying bandgaps and passbands formations; specifically, a decrease in relative density results in wider bandgaps and narrower passbands. Subsequent to the achievements of prior efforts within both design families, researchers were impelled to optimize acoustic metamaterials aimed at audible noise attenuation. Both design families were engaged to optimize the acoustic metamaterials' design, facilitating the exhibition of numerous bandgaps within the audible frequency spectrum (i.e., 20 Hz-20 kHz) [21,22]. An exhaustive search was the optimization technique employed for both studies, which examined hundreds and tens of thousands of designs to identify optimal solutions. The optimal acoustic metamaterial based of the first design family was able to demonstrate bandgaps encompassing 95% of the audible frequency spectrum, whereas the optimal configuration from the second design family exhibited bandgaps covering 85% of the audible frequency spectrum. It is noteworthy that the previously mentioned acoustic metamaterial designs exhibit symmetry. Specifically, the periodicity along the x-axis is equivalent to the periodicity along the y-axis. Consequently, wave propagation remains identical along both periodicities, as the bandgaps and passbands are consistent for each periodicity. However, this symmetry does not benefit most wave guiding applications, which necessitate directional wave guiding. In such cases, distinct bandgaps and passbands must be formed in different periodic directions. Two methodological approaches can achieve distinct bandgaps and passbands along different periodicities. The first approach involves using a demultiplexer, which directs the input signal (i.e., acoustic waves) into two or more channels, resulting in varied propagation. To achieve different wave propagation characteristics across the channels, each channel's geometry should differ. Demultiplexers have been utilized to guide acoustic waves, with some efforts successfully developing Helmholtz resonator-based channels that produce distinct signals from each channel, thus achieving directional wave guiding devices [23,24]. By integrating demultiplexing science with acoustic metamaterials, researchers have developed a demultiplexer acoustic metamaterial [25]. This device contains channels filled with symmetric acoustic metamaterials, each possessing distinct bandgaps and passbands. This configuration allows for different signals to emanate from each channel. In the aforementioned research, a deterministic optimization technique was employed to develop a demultiplexer acoustic metamaterial comprised of two channels capable of exhibiting bandgaps and passbands in the normalized frequency ranges of approximately 0.4 to 0.6 and 0.85 to 1.5. The optimized device functions such that the first channel contains bandgaps in the first desired frequency range and passbands in the second frequency range, while the second channel performs the reverse (i.e., with bandgaps in the second desired frequency range and passbands in the first frequency range). The second approach entails the use of asymmetrical acoustic metamaterials, commonly known as anisotropic acoustic metamaterials. These anisotropic metamaterials are characterized by their asymmetric design, which results in directional bandgaps and passbands along different periodicity axes. Theoretically, this arises from presenting different wave numbers in each periodicity along the 1$^{st}$ Brillouin zone [26]. Some researchers have developed asymmetrical designs of acoustic metamaterials that present different bandgaps and passbands along two different periodicity axes (e.g., periodicities along the x and y axes) [27,28]. Other researchers have started from a symmetrical design of acoustic metamaterials and then applied longitudinal and lateral strains, leading to an asymmetrical shape in the deformed state due to the Poisson's ratio not equaling -1 [29]. These efforts have demonstrated anisotropic acoustic metamaterials which show minimal differences between the bandgaps and passbands along different periodicities. In other words, the periodicities display significant commonalities in terms of bandgaps and passbands. Also, it's great to highlight that another use of anisotropic acoustic metamaterials is to have their refractive index tuned for imaging and detection applications [30,31].

The preceding discussion has demonstrated that acoustic metamaterials represent promising candidates for applications in directional wave guiding. However, the subsequent inquiry pertains to which metamaterial is more feasible for such applications: the demultiplexer or the anisotropic acoustic metamaterial. Both are great candidates, however, the demultiplexer is more geared towards specific wave guiding applications that deal with channel-based systems. The anisotropic metamaterial emerges as a more versatile choice, as it relies on a single metamaterial system that could be employed without channels, thereby rendering them as preferred candidates for most of the directional wave guiding applications. Furthermore, from a design perspective, it is less computationally expensive since the designer will only be required to optimize a single metamaterial system. Nonetheless, an issue persists with current anisotropic acoustic metamaterial designs, as noted earlier, which is the occurrence of commonalities in bandgaps and passbands across different periodicities, which impede their utility for directional wave guiding applications. To tackle this issue, new anisotropic metamaterials designs should be discovered. Generally, the common framework in discovering new designs in engineering is geometrically modifying pre-existing designs. One of the most common approaches for creating new 3-dimensional designs is the hybridization of two existing designs. This design approach was used to create hybrid TPMS-based lattices which demonstrated superior mechanical and thermal properties [32–37]. Motivated by the aforementioned literature, this work's objective is to utilize the hybridization design approach for creating new anisotropic acoustic metamaterial featuring distinctly unique bandgaps and passbands across different periodicities. This research will employ optimization technique (i.e., exhaustive search) to refine anisotropic metamaterial designs suitable for functioning as excellent passive directional wave guiding devices within specific frequency ranges.

## 2. Description

### 2.1 Unit Cell Design

The acoustic metamaterial is characterized as a periodic structure composed of a recurrent fundamental element that defines its periodicity. This fundamental element is termed as the unit cell. As a result, the topology of the acoustic metamaterial is contingent upon the design of the unit cell. The design proposed in this study is termed as hybrid unit cells, with their depictions provided in Figure 1. This proposed design of the unit cell is achieved through the hybridization of two distinct unit cells (referred to in this work as subunit cells). The selected designs for the subunit cells are the sheet-based version of TPMSs. Four TPMSs have been considered: the Primitive, Gyroid, Diamond, and Nevious. The designs examined in this research comprise two subunit cells constructed from the same TPMS structure but differing in their geometric parameters (i.e., also referred to as design variables). Consequently, this study will encompass four distinct design families: Primitive-Primitive, Gyroid-Gyroid, Diamond-Diamond, and Nevious-Nevious. Figure 2 illustrates the design families considered.

$$X = \left[\rho_1^*, \rho_2^*, \frac{C_2}{C_1}, C_1\right]^T \tag{1}$$

Equation 1 presents the design variables vector. Where $X$ denotes to the design variables. Where $\rho_1^*, \rho_2^*$ denote to the relative densities of the two subunit cells. It is important to highlight that the relative densities for these structures are found through curve-fitted based functions which can be found in the previous literature [22,38]. $C_1, C_2$ are the sizes of the subunit cells. The ratio between the subunit cells ($C_2/C_1$) is considered as design variable.

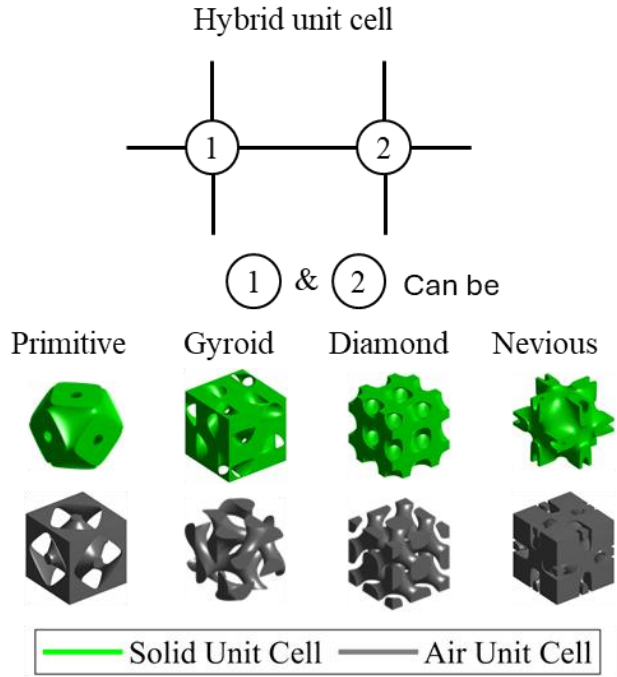

Figure 1 The hybrid metamaterial designs.

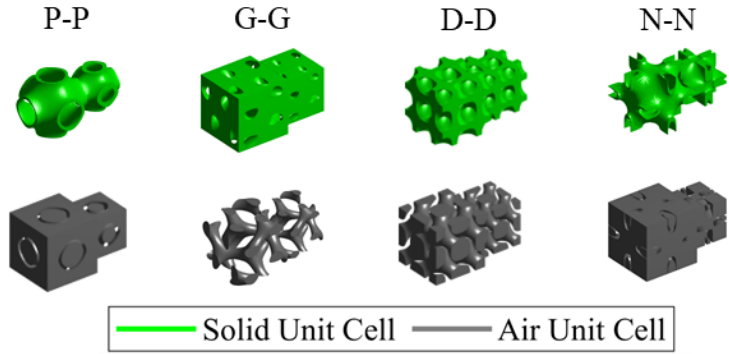

Figure 2 The design families considered. P-P,G-G,D-D, N-N denote to Primitive-Primitive, Gyroid-Gyroid, Diamond-Diamond and Nevious-Nevious.

## 2.2 Design Advantages

The proposed metamaterial designs, referred to as hybrid metamaterials, exhibit asymmetrical characteristics due to differing periodicity patterns along the x and y axes. The x-axis features a semi-periodic pattern, whereas the y-axis exhibits a fully periodic pattern. The design variables considered in this study result in two design scenarios for the proposed hybrid metamaterials, specifically concerning the ratio between the size of the second subunit cell and the first subunit cell ($C_2/C_1$). Figure 3 illustrates these two design scenarios. In the first scenario, where $C_2/C_1$ equals 1, both subunit cells (i.e., first and second subunit cells) in the metamaterial are connected to the adjacent unit cell along y-periodicity, as shown in Figure 3-A. Conversely, the second scenario, where $C_2/C_1$ is less than 1, results in only the first subunit cell being connected to the adjacent unit cell along the y-periodicity, as depicted in Figure 3-B.

The second design scenario has been identified to result in metamaterial constructs that are optimally configured for directional waveguiding applications. The discontinuous interfaces between the adjacent second subunit cells along the y-periodicity within the metamaterial are engineered to adhere to rigid boundary conditions, as depicted in Figure 4. This illustration delineates the solid part of the metamaterial that complies with the prescribed boundary conditions imposed on its air part. As illustrated, the faces of the second subunit cell aligned with the y-periodicity are obstructed by a solid barrier, thereby inhibiting the transmission of sound waves along this subunit cell to its neighboring counterparts. Consequently, the sole feasible pathway for sound propagation through this periodicity is via the interface connectivity between adjoining first subunit cells. This characteristic of sound transmission confers the design with bandgap-rich features along the x-periodicity and passband-rich attributes along the y-periodicity, culminating in a pronounced reduction of bandgap commonalities across both periodicities. Figure 5 exemplifies a metamaterial adhering to this design paradigm, presenting its unit cells (illustrated in Figure 5-A) and respective bandgap-structures along both periodicities (as shown in Figures 5-B and 5-C). The black regions denote the bandgaps, whereas the white regions indicate the passbands within the bandgap-structure. The x-axis in the bandgap-structure represents the wavenumbers denoting the periodicity (i.e., referred to as directional wave number in Figures 5-B and 5-C).

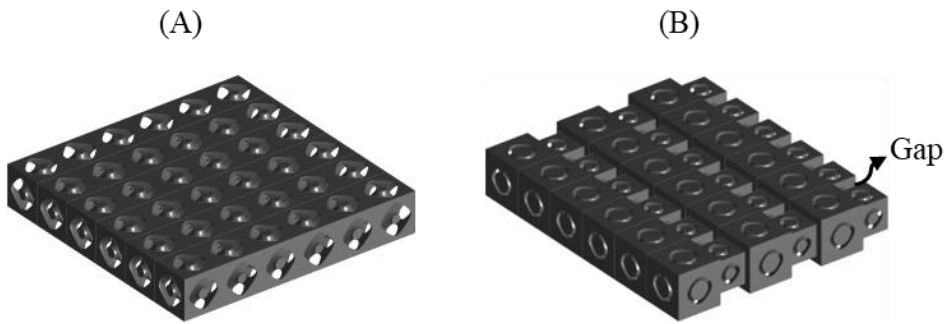

Figure 3 The hybrid metamaterials resulted from the two scenarios. A) The metamaterial resulted from the first scenario. B) The metamaterial resulted from the second scenario.

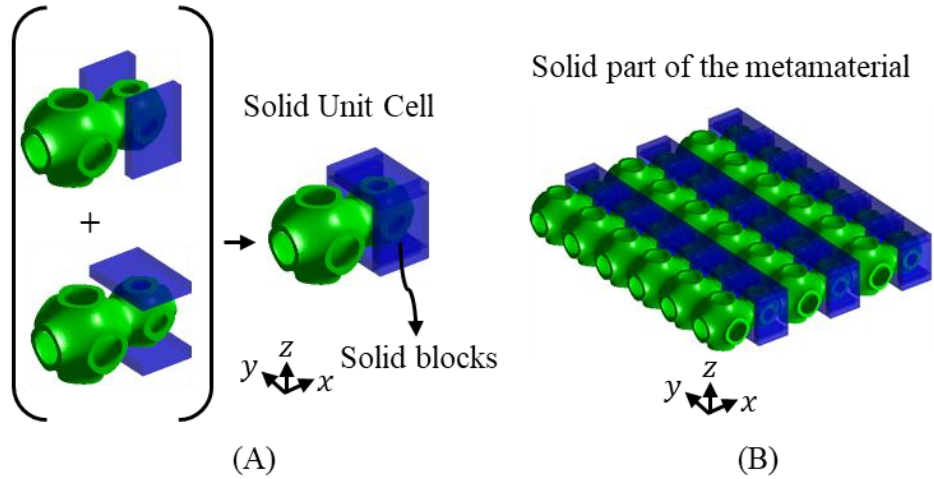

Figure 4 The solid part of the metamaterial which follows the conditions for the second design scenario. A) The solid unit cell. B) The solid part of the metamaterial.

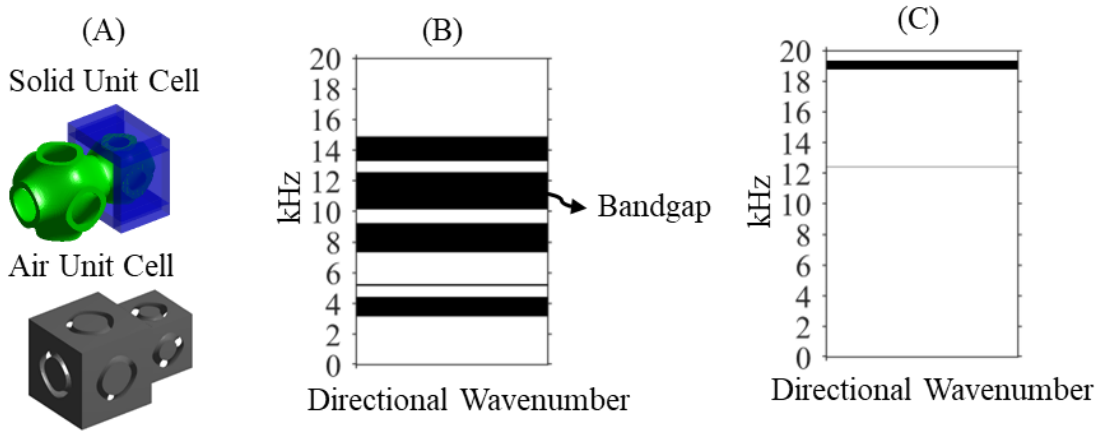

Figure 5 A demonstration of the efficacy of second design scenario for developing metamaterial with extreme bandgaps and passbands characteristics. A) The air unit cell of the metamaterial. B) Bandgaps-structure along the x-periodicity. C) Bandgaps-structure along the y-periodicity.

## 2.3 Bandgaps and Passbands Computations

The bandgaps and passbands of the hybrid metamaterials are computed numerically through utilizing Finite Element (FE)-Bloch's theorem methodology. This method utilizes the eigen-value problem of the equation governing wave propagation along with the incorporation of Bloch's theorem. Through the application of Bloch's theorem, the unit cell of the periodic metamaterial is sufficient enough to be the geometrical domain for capturing the bandgaps and passbands of the whole structure.

$$(-\omega^2 M + K)[P] = 0 \qquad (2)$$

Equation 2 presents the FE discretized eigen-value problem presenting the sound wave propagation. This equation is the FE discretized eigen-value problem belonging to the Helmholtz equation (i.e., Partial Differential Equation that presents the sound wave propagation) [39]. Where $P$ denotes to the acoustic sound pressure, $\omega$ are the eigen frequencies. $M, K$ are the mass and stiffness matrices. Impedance mismatch is assumed, which states that when there are two mediums existing in the system with large differences in impedances (i.e., one medium has much higher impedance than the other) the sound will propagate through the medium having lower impedance. Subsequent to the incorporation of such assumption, the air medium is the considered domain in which sound propagates, with physical properties of density $\rho = 1.2\ kg/m^3$ and speed of sound $c = 343\ m/s$. Therefore, the air unit cell is the geometrical domain considered in the FE-Bloch's theorem methodology. An illustration of the air unit cell is depicted in Figure 2.

In order to utilize the air unit cell as the geometrical domain that is able to capture the bandgaps and passbands of the corresponding periodic metamaterial, Bloch's theorem had to be applied. For this theorem to be satisfied, Periodic Boundary Conditions (P.B.Cs) should be applied on the tessellated faces. Figure 6 illustrates the application of the P.B.Cs on the tessellated faces.

$$P_2(\omega) = P_1(\omega)e^{-ik.r_1}; r_1 = C_x i + 0j + 0k \qquad (3.1)$$

$$P_4(\omega) = P_3(\omega)e^{ik.r_2}; r_2 = 0i + C_y j + 0k \qquad (3.2)$$

Equation 3 present the P.B.Cs applied on the tessellated face, where equation 3.1 represent the P.B.Cs applied on the tessellated faces along the x-periodicity, while equation 3.2 represent the P.BC.s applied on the tessellated faces along y-periodicity. Here, $r_1, r_2$ denote to the lattice vectors that characterize the periodicity of the metamaterial through the unit cell. In simpler words, the unit cell is translated multiple times using these lattice vectors to form the corresponding periodic metamaterial. $r_1$ presents the lattice vectors characterizing the periodicity along the x-axis while $r_2$ presents the lattice vectors characterizing the periodicity along the y-axis as illustrated in Figure 6-A. $C_x$ and $C_y$ are the unit cell sizes along the x-axis and y-axis. Here $\boldsymbol{k}$ denotes the possible wavenumbers in which the metamaterial propagates with.

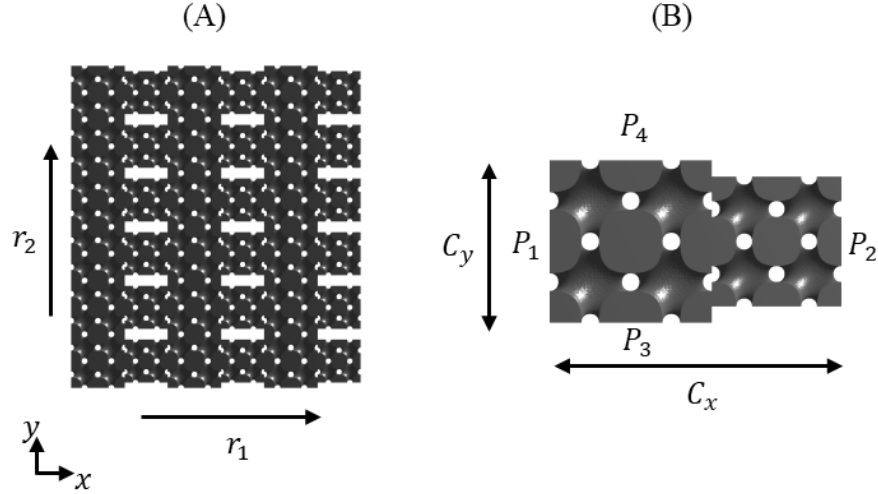

Figure 6 Illustration of the metamaterial along with its corresponding lattice vectors and P.B.Cs. A) Air part of the metamaterial. B) Air unit cell.

The possible wave numbers ($k$) of the unit cells are found through the 1st Brillouin zone [26]. Figure 7 presents the 1st Brillouin zone of the unit cells considered in this work. This zone is established through finding the reciprocal vectors in the half-pi space ($b_1, b_2$). Where $b_1 = \frac{\pi}{C_x} i + 0j + 0k$ and $b_2 = 0i + \frac{\pi}{C_y} j + 0k$. Looking at Figure 7-A, the 1st Brillouin zone for these unit cell designs is a region denoted as $\Gamma - X - M - Y - \Gamma$, which is a region that presents the possible wave numbers of the metamaterial along propagations in all directions. However, this work is only concerned in directional wave propagation, therefore the regions presented Figures 7-B and 7-C are the ones considered. These regions encapsulate the wave numbers in which the metamaterial propagate with along the x and y periodicities. Where $\Gamma - X$ presented in Figure 7-B encapsulates the wave numbers for propagation along x-periodicity and $\Gamma - Y$ presented in Figure 7-C encapsulates the wave numbers for propagation y-periodicity.

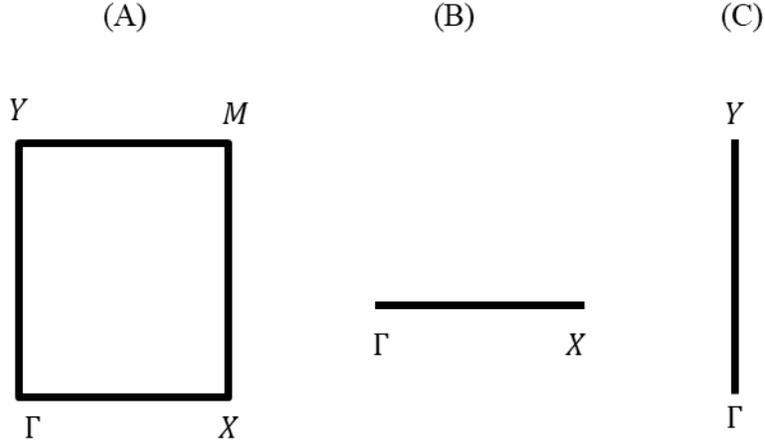

Figure 7 1st Brillouin Zone. A) The full 1st Brillouin zone (i.e., $\Gamma - X - M - Y - \Gamma$). B) The region presenting the wavenumbers for propagation along x-periodicity (i.e., $\Gamma - X$). C) The region presenting the wavenumbers for propagation along y-periodicity (i.e., $\Gamma - Y$).

Upon considering Equations 3.1 and 3.2 along with regions of the 1st Brillouin zone presented in Figures 7-B and 7-C, Equation 2 becomes

$$(-\omega(\boldsymbol{k})^2 M + K)[P_r] = 0 \qquad (4)$$

Equation 4 is the reduced and updated system for Equation 2 subsequent to the incorporation of the P.B.Cs. Where $P_r$ denote to the reduced acoustic sound pressures. This equation is a function of the wavenumbers. Meaning that, for one unit cell design, the equation will be solved for every wavenumber in the regions considered (i.e., $\Gamma - X, \Gamma - Y$) and the eigenfrequencies ($\omega$) are found. Upon finding the eigenfrequencies ($\omega$) for one region (i.e., $\Gamma - X, \Gamma - Y$), a band-structure is formed, where frequency ranges that are covered by the eigenfrequencies refer to the passbands while frequency ranges that aren't covered by the eigenfrequencies refer to bandgaps.

Equation 4 is solved numerically using FE software package ABAQUS. FREQUENCY STEP was utilized, particularly, AMS SOLVER was considered. EQUATION function was used to incorporate the P.B. Cs listed in Equations 3.1 and 3.2 in the numerical simulation. AC3D10 is the element used to discretize the air unit cell into 3-dimensional quadratic elements comprised of 10 nodes. Mesh sensitivity was conducted and found that the results converge with the maximum and minimum element sizing of 3 and 0.5 mm.

It is great to highlight that the bandgaps and passbands are stored in their normalized format.

$$n = \frac{fC_1}{c} \tag{5}$$

Equation 5 presents the normalization considered in this work. Where $n$ refers to the normalized eigenfrequency, $f$ is the actual eigenfrequency in Hz. As shown in this equation, the normalization is done with respect to the size of the first subunit cell ($C_1$).

## 2.4 Design Objectives and Framework

The principal aim of this research is to identify optimal designs for hybrid metamaterials that can be effectively utilized for directional wave guiding applications across various frequency domains. The frequency ranges explored in this study are confined to the audible spectrum, specifically from 20 Hz to 20 kHz. Table 1 enumerates the frequency ranges considered. The process of design optimization for the hybrid metamaterials will be undertaken for each frequency range specified. The optimal hybrid metamaterial design would be one that demonstrates bandgaps along the x-periodicity, covering the majority of the targeted frequency range, and passbands along the y-periodicity, covering the same targeted frequency range.

$$O.F = \min[B.C_x, P.C_y] \tag{6}$$

Equation 6 presents the objective function considered in this work. Where $O.F$ denotes to the objective function, $B.C_x$ and $P.C_y$ are the bandgaps and passbands coverage along both periodicities in the desired frequency range. As presented in this equation, the optimization technique will maximize the minimum from both bandgaps and passbands along both periodicities in the desired frequency range. This objective function was shown to be more effective than simply taking the magnitude of both coverages (i.e., $B.C_x$ and $P.C_y$) since it ensures the existence of sufficient $B.C_x$ and $P.C_y$ in the desired frequency range. In other words, the consideration of maximizing the magnitude of both coverages may lead to an optimal design which has a high $B.C_x$ but a low $P.C_y$. Figure 8 illustrates the objective of optimization in this work.

Table 1 The desired frequency ranges considered in this work.

| Case | Desired Frequency Range |
|------|------------------------|
| 1 | 20 Hz-5 kHz |
| 2 | 5 kHz-10 kHz |
| 3 | 10 kHz-15 kHz |
| 4 | 15 kHz-20 kHz |

The optimization framework employed in this study incorporates a computationally efficient exhaustive search method, which has been previously utilized in our earlier efforts [21,22]. Figure 9 illustrates the exhaustive search framework applied for the optimization of the designs examined in this work. This framework can be broadly delineated into two major steps.

The first major step entails the generation of normalized designs facilitated by extensive parametric study. This phase is subdivided into several sub-steps. First, the normalized design variables (i.e., $\rho_1^*, \rho_2^*, C_2/C_1$) are subject to variation. Second, normalized design combinations are generated via an in-house computational framework that employs MATLAB scripts, Python scripts, and FreeCAD. Third, a design constraint is incorporated to ensure the selection of only those designs deemed logical. Logical designs, for the purposes of this study, are characterized by face interfaces between the two subunit cells. Figure 10 delineates the logical designs (refer to Figure 10-A) and non-logical designs (refer to Figure 10-B). Fourth, following the attainment of a set of logical designs, the bandgaps are computed employing the FE-Bloch's theorem, as elaborated in the preceding subsection. It is imperative to recall and note that the bandgaps are retained in their normalized format.

The second step entails converting these normalized bandgap data into actual ones and determining the optimal unit cell design. This process is established through several sub-steps. First, a set of first subunit cell sizes ($C_1$) is considered. Second, this set, composed of different $C_1$ sizes, is integrated into the normalized bandgaps using Equation 5 to generate a series of actual designs along with their bandgap data. It is noteworthy that each normalized design yield multiple actual counterparts, contingent upon the size of the $C_1$ set. The $C_1$ set considered in this work comprises 401 distinct sizes, ranging from 10 mm to 50 mm, with increments of 0.1 mm. Therefore, for this study, each normalized design corresponds to 401 actual counterparts with different $C_1$ sizes. As previously documented in earlier research [21,22], this method is computationally efficient. Rather than employing FE-Bloch's theorem to compute the actual bandgaps, these bandgaps are derived through a straightforward arithmetic operation. Third, a design constraint is applied, which considers only actual designs where the cavities are 20 times larger than the entropy boundary layer thickness. Detailed elaboration on this step is available elsewhere [22]. Fourth, the objective function is calculated for each design, facilitating the identification of the optimal design for each of the assessed frequency ranges.

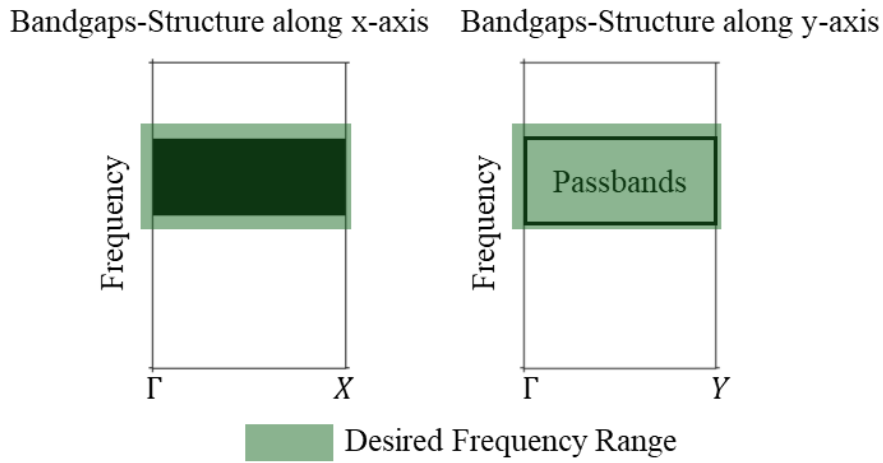

Figure 8 Illustration of this wok objective.

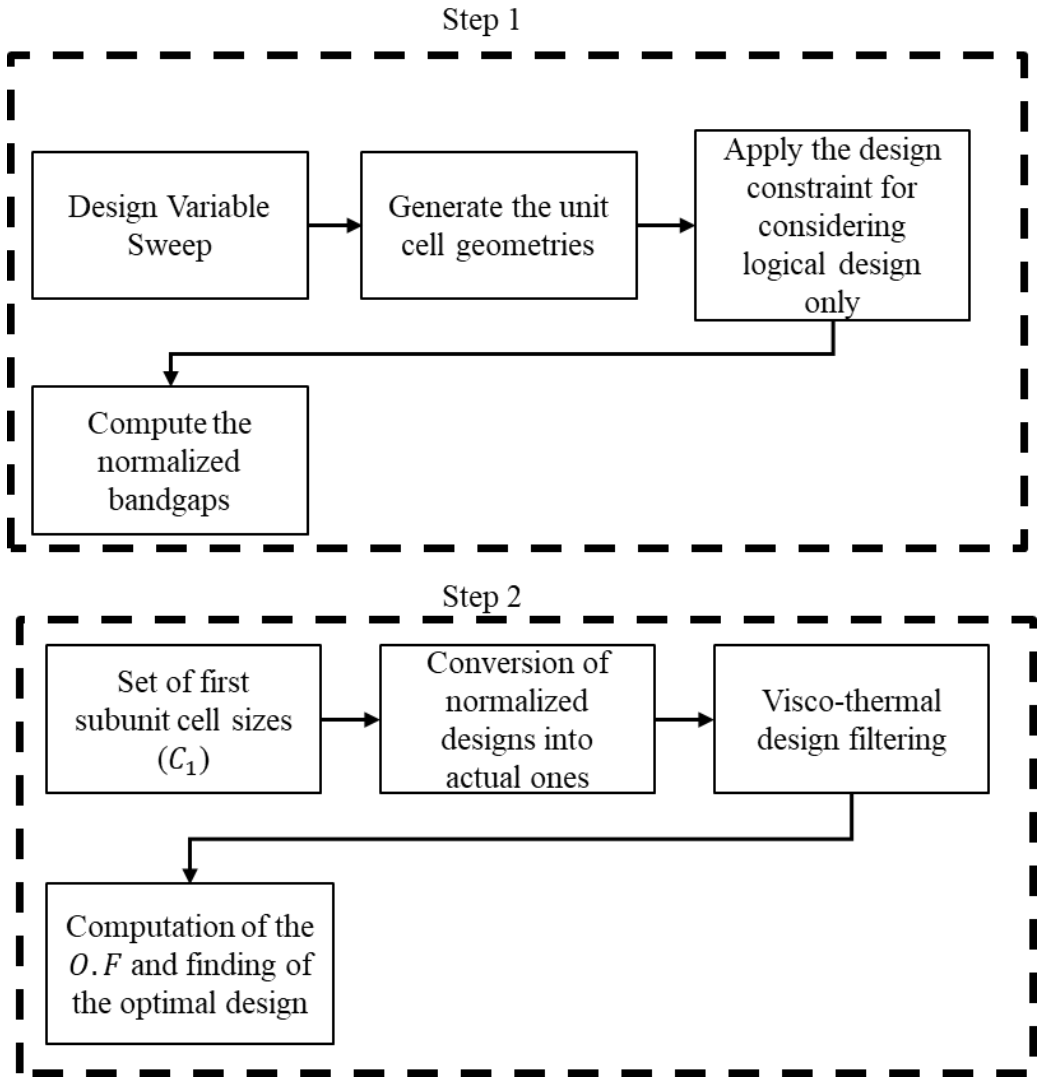

Figure 9 Exhaustive search steps.

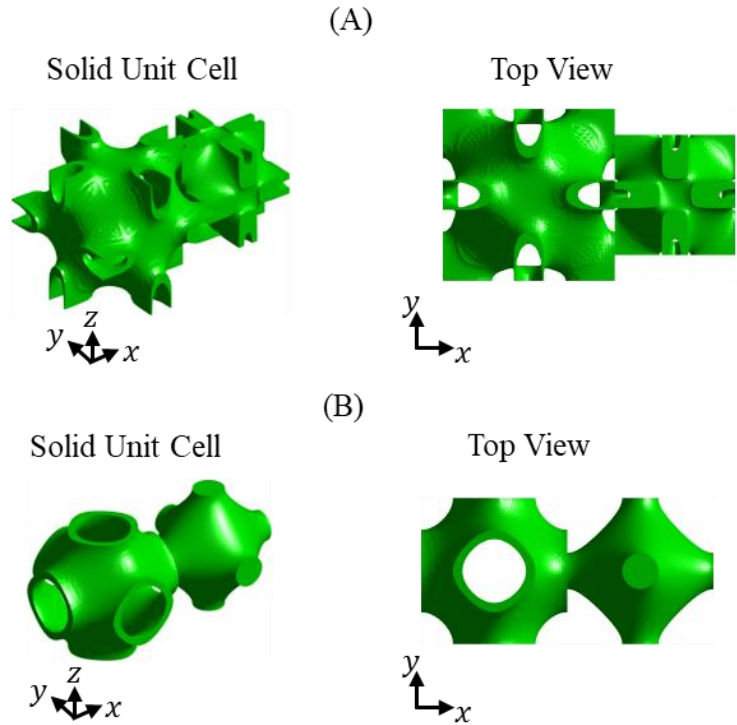

Figure 10 Illustration of the logical and non-logical designs. A) Logical designs. B) Non-logical designs.

## 3. Results and Discussion

### 3.1 Optimal Unit Cell Designs from the Exhaustive Search

There are two designs spaces considered in this work: the normalized and the actual design spaces. Figure 11-A illustrates the normalized designs for each of the design family. The depicted normalized designs are found subsequent to applying the design constraint which filters in the logical designs. A total of 978 normalized designs were considered. Figure 11-B shows the actual designs considered for each design family. A total of 392178 designs were considered for finding the optimal design in each frequency range.

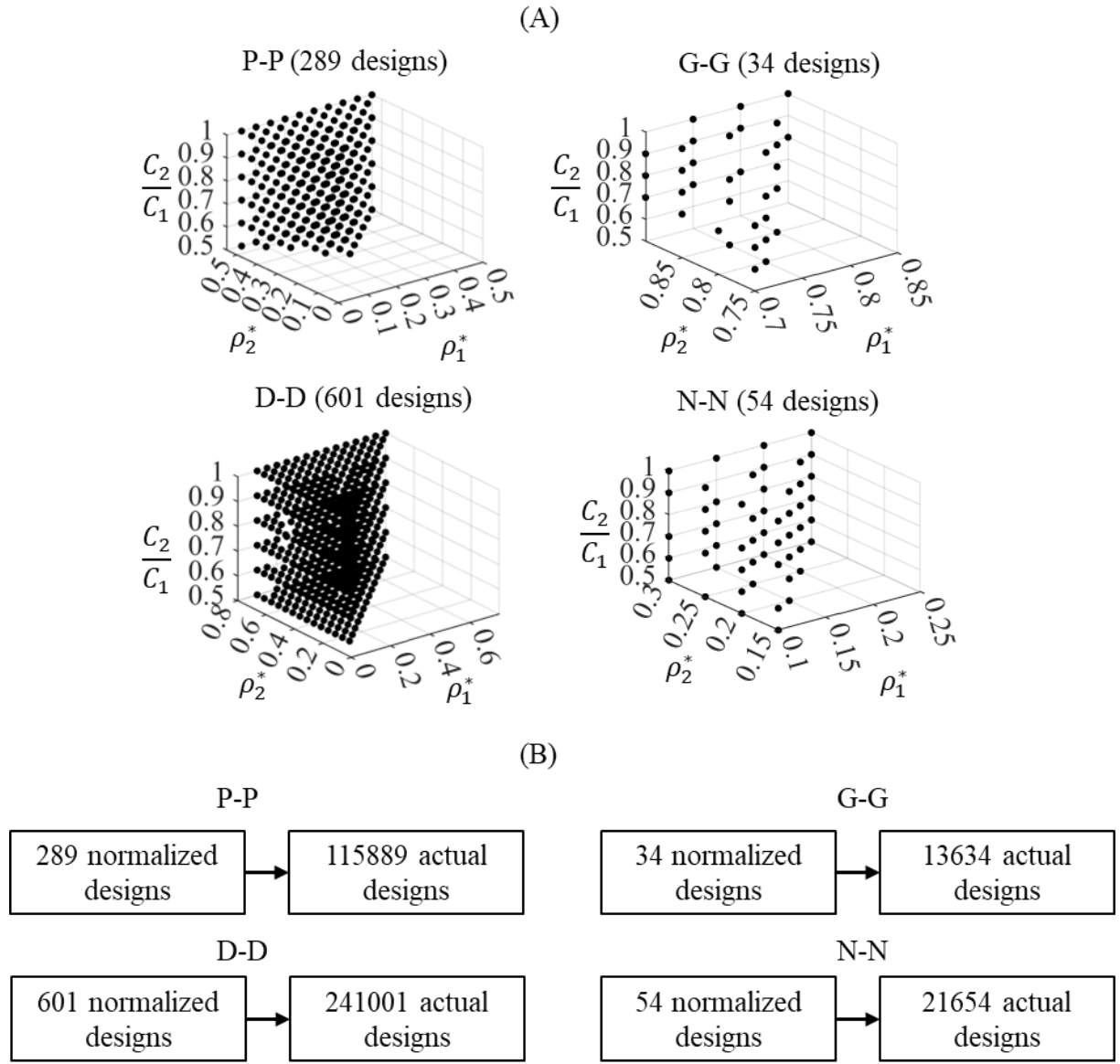

Figure 11 The designs considered in the exhaustive search steps. A) Normalized designs presented as points. B) The conversion from normalized designs into their actual counterparts.

Figure 12 illustrates the optimal unit designs in conjunction with their respective bandgap structures. Table 2 enumerates the design variables and the objective function values of the optimal designs for each frequency range. Upon examining the bandgap structures of the optimal designs across all frequency ranges, it is evident that these hybrid designs manifest significant anisotropic effects. The objective of this study, which entailed the optimization of bandgaps along the x-periodicity and passbands along the y-periodicity for various audible frequency ranges, was realized through these designs. The observed bandgap and passband behavior across both periodicities represents a significant achievement not previously documented in the literature. Furthermore, the anisotropic effect of such designs becomes more pronounced at higher audible frequency ranges. For example, the optimal design targeting the frequency range of 15 kHz to 20 kHz, which belonged to the Diamond-Diamond sheet-based design family with the design variables of ($\frac{C_2}{C_1} = 1, \rho_1^* = 20, \rho_2^* = 80\%, C_1 = 48.2\ mm$), achieved an objective function value of 93.53%. However, the most important optimal design would be the optimal one for targeting the lowest audible frequency range (i.e., 20 Hz to 5 kHz), which belonged to the Nevious-Nevious design family with the design variables of ($\frac{C_2}{C_1} = 0.7, \rho_1^* = 10\%, \rho_2^* = 30\%, C_1 = 50\ mm$) and had an objective function value of 62.90%. The rationale behind favoring this optimal design among the others is due to the fact that most of the audible noises pass through the frequency range of 20 Hz to 5 kHz.

Table 2 The optimal unit cell design variables and objective functions values.

| Design Family | $C_2/C_1$ | $\rho_1^*$ (%) | $\rho_2^*$ (%) | $C_1$ (mm) | $O.F$ (%) |
|---|---|---|---|---|---|
| Nevious-Nevious sheet-based | 0.7 | 10 | 30 | 50 | 62.90 |
| Gyroid-Gyroid sheet-based | 1 | 80 | 85 | 42.9 | 76.71 |
| Diamond-Diamond sheet-based | 0.7 | 40 | 70 | 48.8 | 83.35 |
| Diamond-Diamond sheet-based | 1 | 20 | 80 | 48.2 | 93.53 |

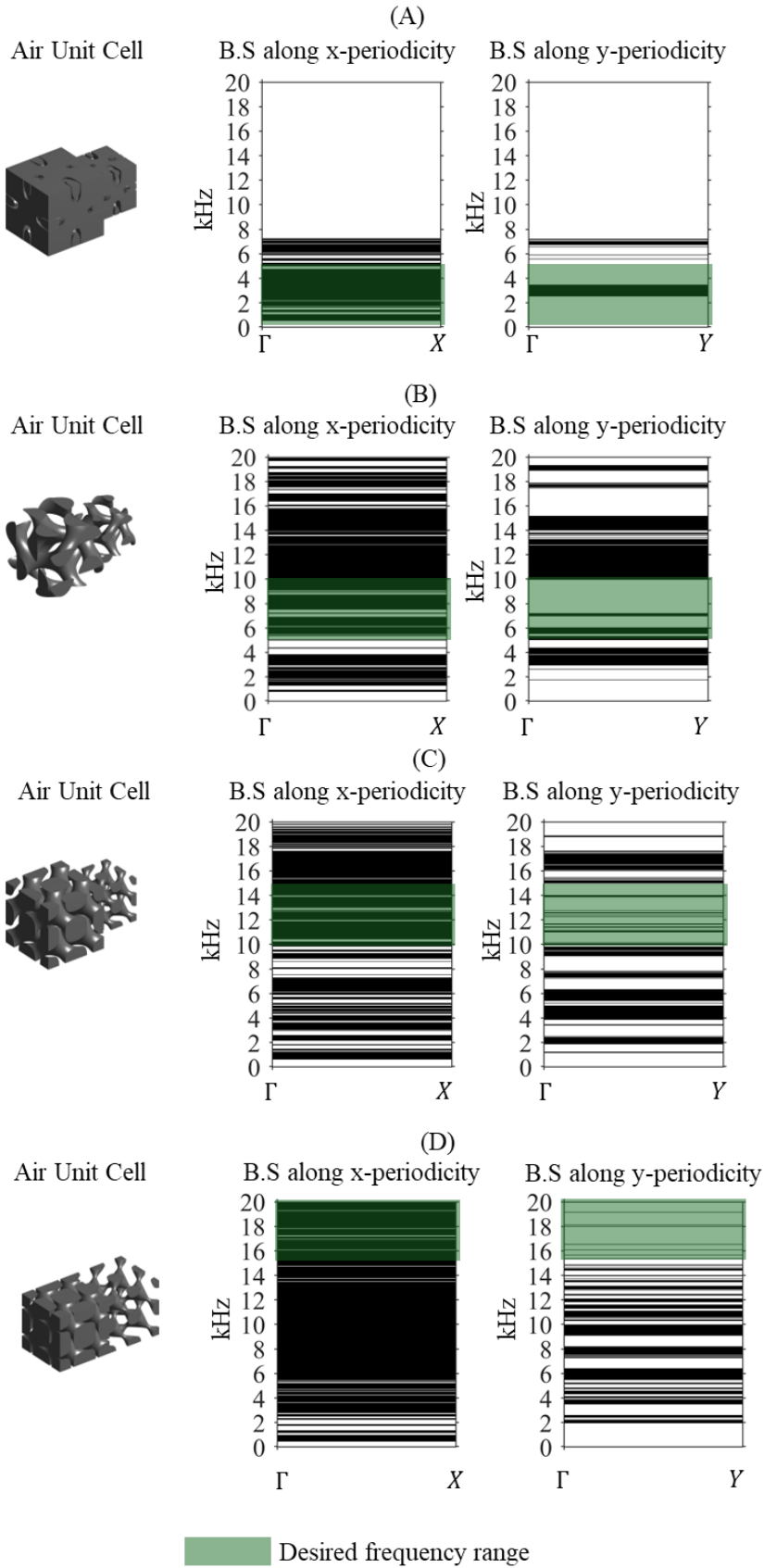

Figure 12 The optimal designs along with their corresponding bandgaps-structures (B.S) for the considered frequency ranges. A) Optimal design for the frequency range of 20 Hz to 5 kHz. B) Optimal design for the frequency range of 5 kHz to 10 kHz. C) Optimal design for the frequency range of 10 kHz to 15 kHz. D) Optimal design for the frequency range of 15 kHz to 20 kHz.

### 3.2 The Acoustic Pressure Response of Hybrid Metamaterials

To further verify and validate the existence of these directional bandgaps and passbands in the optimal designs, the Acoustic Pressure Responses (APRs) of hybrid metamaterials made from repeated optimal unit cells were computed. The APRs were numerically computed using FE method. The direct solver was utilized to compute the APRs in the FE simulations.

$$P(\omega) = [-\omega^2 M + K]^{-1} F(\omega) \qquad (7)$$

Equation 7 presents the direct solver equation. This equation is a set of coupled linear algebraic equations which is solved for finding the acoustic sound pressure at the discretized unknown nodes. It is great to highlight that this equation is derived subsequent to the FE-discretization and Laplace transform of the Helmholtz equation. Here, $P$ refers to the acoustic sound pressure at the unknown nodes, $F$ is the external sound pressure applied and $\omega$ is the driving frequency. ABAQUS is utilized to solve the equation. Particularly, STEADY STATE DYNAMICS STEP is utilized. Same element type and mesh sensitivity were considered in this analysis as the ones for the bandgaps' computations.

The hybrid metamaterial chosen for this analysis is the metamaterial made from repeated optimal Nevious-Nevious sheet-based unit cell with the design variables of ($\frac{C_2}{C_1} = 0.7, \rho_1^* = 10\%, \rho_2^* = 30\%, C_1 = 50\ mm$). As mentioned previously, this unit cell was found to be the optimal design for targeting the frequency range of 20 Hz to 5 kHz. Figure 13 presents the air part of the selected hybrid metamaterial along with the actuation scenarios and sensing locations considered. A hybrid metamaterial made from 3 by 3 repeated unit cell was considered as illustrated Figure 13-A. Two actuation scenarios were considered and are presented in Figures 13-B and 13-C. The first actuation scenario (i.e., illustrated in Figure 13-B and referred to as Actuation Scenario I) involves exciting the system with an external sound pressure at the face located at one end of the metamaterial along the x-axis, this makes the wave propagate along x-periodicity. While the second actuation scenario (i.e., illustrated in Figure 13-C and referred to as Actuation Scenario II) involved exciting the system with an external sound pressure at the face located at one end of the metamaterial along the y-axis, this makes the wave propagate along y-periodicity. The same two sensing locations were considered for both actuation scenarios (i.e., denoted as X and Y), which were points located at the other ends of the metamaterial along the x and y axes as presented in Figures 13-B and 13-C. The external sound pressure was in the form of chirp signal with the magnitude of $P_i = 1 Pa$ and applied on the frequency range of 20 Hz to 5 kHz.

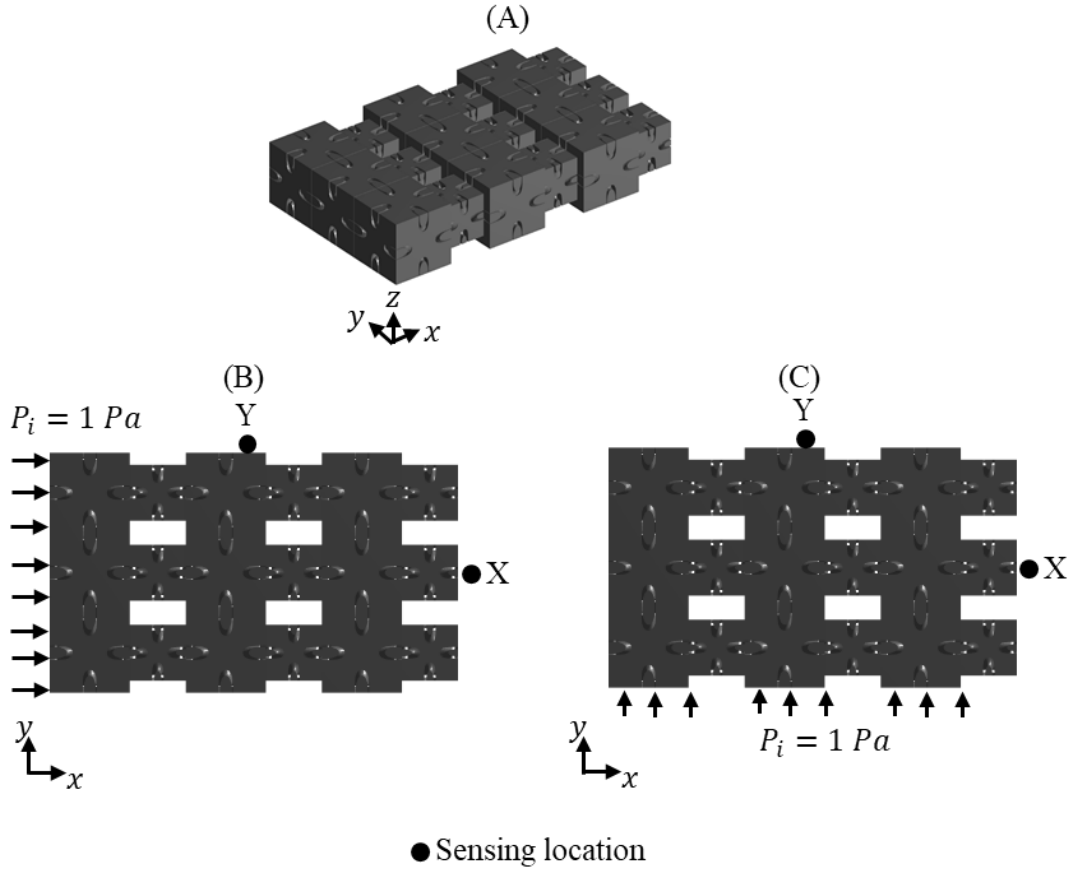

Figure 13 The hybrid metamaterial selected along with the actuation scenarios and sensing locations considered. A) Illustration of the air part of the metamaterial. B) Illustration of Actuation Scenario I. C) Illustration of Actuation Scenario II.

Figure 14 shows the numerically computed APRs at the considered sensing locations for both actuation scenarios. This figure is organized into two subplots, each subplot represents the APRs at one sensing location for each scenario (i.e., Figure 14-A presenting APRs at sensing location X and Figure 14-B presenting APRs at sensing location Y). Subsequent to the observations of Figures 14-A and 14-B, both sensing locations X and Y had shown attenuation in frequency ranges corresponding to the bandgaps along x-periodicity when Actuation Scenario I was considered, while the same sensing locations had shown attenuation in frequency ranges corresponding to the bandgaps along y-periodicity when Actuation Scenario II was considered (refer to Figure 12-A). It is great to highlight that the sound pressure levels ($SPLs$) for both sensing locations at Actuation II is noticeably higher than ones for Actuation Scenario I. The aforementioned observations show that this hybrid metamaterial is efficient in directing the waves along both periodicities for its targeted frequency range (i.e., 20 Hz to 5 kHz).

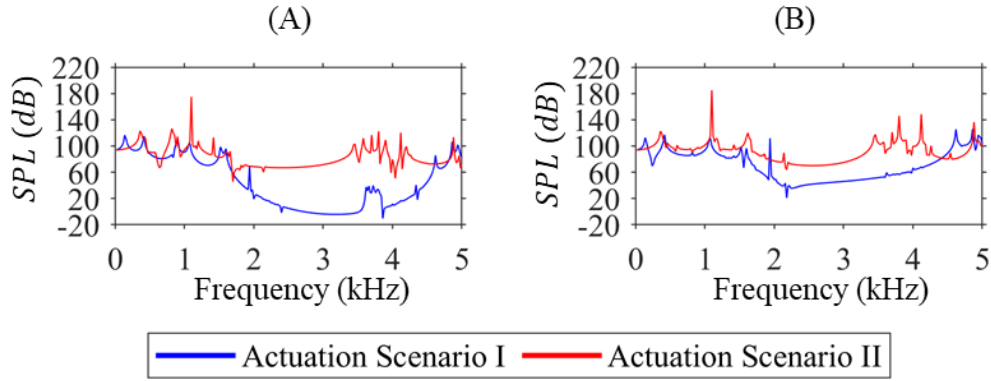

Figure 14 The numerically computed APRs. A) APRs at sensing location X. B) APRs at sensing location Y.

## 4. Conclusion

The primary objective of this study was to develop effective anisotropic acoustic metamaterials for applications in directional wave guiding. To achieve such effectiveness, the study sought to design metamaterials with extreme bandgap and passband characteristics along x and y periodicities. To attain this level of bandgap and passband exhibition, a new class of designs, termed hybrid metamaterials, was proposed. These designs consist of unit cells that hybridize two subunit cells with different design variables to form the metamaterial. Four design families were considered in this study (i.e., Primitive-Primitive, Gyroid-Gyroid, Diamond-Diamond, and Nevious-Nevious). These selected designs exhibited extreme bandgap and passband characteristics. Specifically, hybrid unit cell designs with $C_2/C_1$ less than 1 demonstrated bandgap-rich characteristics along the x-periodicity and passband-rich characteristics along the y-periodicity, rendering them viable design candidates for directional wave guiding applications. Following this realization, these hybrid designs were optimized to exhibit bandgaps and passbands across four desired frequency ranges (i.e., 20 Hz to 5 kHz, 5 kHz to 10 kHz, 10 kHz to 15 kHz, and 15 kHz to 20 kHz). The optimization technique employed was exhaustive search, which involved two major steps. The first step entailed conducting an extensive parametric analysis that included 978 normalized designs and calculated their bandgaps. The second step utilized these normalized designs and their bandgap data to convert them into 392,178 actual design combinations. Subsequently, objective functions were computed, and optimal designs were identified for each of the specified frequency ranges. It was demonstrated that the optimal unit cell designs exhibited superior bandgap and passband characteristics. Furthermore, as the frequency range of interest went higher, the corresponding optimal design exhibited a higher objective function. For instance, the optimal unit cell design targeting the frequency range of 15 kHz to 20 kHz had an objective function value of 93.53%. However, the most advantageous optimal unit design would be the one targeting the lowest frequency range (i.e., 20 Hz to 5 kHz). This is the range where most of the noise occurs, thus making it imperative to consider the optimal design for this range. For cross-verification of the existence of bandgaps and passbands in the optimal metamaterials, the APRs were computed for two actuation scenarios, alongside the consideration of two sensing locations. The computed APRs demonstrated results that showed a correspondence between their attenuation regions and the bandgaps along both periodicities. Specifically, the APRs at both sensing locations for Actuation Scenario I agreed with the bandgap structure along x-periodicity, while those for Actuation Scenario II agreed with the bandgap structure along y-periodicity.

**Acknowledgements**

This research received funding from both Natural Sciences and Engineering Research Council of Canada through Dr. Salehian's Discovery Program as well as the authors' funding with the Government of Canada through the Federal Economic Development Agency for Southern Ontario (FedDev Ontario) in the Waterloo Institute for Sustainable Aeronautics at the University of Waterloo. This work was partially conducted using the High-Performance Computing devices through the use of Digital Alliance of Canada services.

**Conflict of Interest**

The authors have no conflicts to disclose.

**Data Availability**

The data that support the findings of this study are available from the corresponding author upon reasonable request.